\documentclass[12pt,preprint]{aastex}
\usepackage{natbib}
\usepackage{rotating}

\shortauthors{Marino et al.}
\shorttitle{Tracing rejuvenation events in nearby S0 galaxies}

\begin{document}

\title{Tracing rejuvenation events in nearby S0 galaxies}

\author{
Antonietta Marino\altaffilmark{1},   
Luciana Bianchi\altaffilmark{1}, 
Roberto Rampazzo\altaffilmark{2}, 
David A. Thilker\altaffilmark{1},
Francesca Annibali\altaffilmark{2},
Alessandro Bressan\altaffilmark{2},
Lucio Maria Buson\altaffilmark{2}
}
\altaffiltext{1}{ 
Dept. of Physics \& Astronomy, Johns Hopkins
University, 3400 N. Charles St., Baltimore, MD 21218;
amarino@pha.jhu.edu}
\altaffiltext{2}{INAF Osservatorio Astronomico di Padova, Vicolo dell'Osservatorio~5, I-35122  Padova,
Italy}
 
\begin{abstract}

With the aim of characterizing rejuvenation processes in early-type galaxies,
we analyzed five barred S0 galaxies showing prominent outer ring  
 in ultraviolet (UV) imaging.  We analyzed  GALEX far- (FUV)  and near- (NUV)
UV and optical data using stellar population models and
estimated the age and the stellar mass of the entire galaxies  and of the  UV-bright  ring
structures.   Outer rings  consist of  young ($\lesssim$200 Myr old) stellar 
populations, accounting for up to 70\% of the FUV flux but containing
only  a few \% of the total stellar mass. 
Integrated photometry of the whole galaxies places four of these
objects on the green valley,  indicating a globally evolving nature. We
suggest such galaxy evolution is likely driven by bar induced 
instabilities, i.e. inner secular evolution, that conveys gas to the nucleus
and  to the outer rings. At the same time, HI observations  of NGC 1533 and 
NGC 2962 suggest external gas re-fueling can play a role in the 
rejuvenation processes of such galaxies.

\end{abstract}


\section{Introduction}
\label{intro}

A large spread in luminosity-weighted ages of early-type galaxies 
(ETGs hereafter), derived by  modeling spectral line-strength indices, 
suggests that {\rm the nucleus} may have been rejuvenated by 
recent, secondary star formation episodes 
\citep[Äe.g.][and references therein]{Annibali07}. 
Spitzer  nuclear spectra of ETGs support this view. 
Polycyclic Aromatic Hydrocarbons emission, characterizing  both
starburst and post-starburst phases, are detected in a large fraction of ETG
nuclear spectra \citep[e.g.][and references therein]{Panuzzo10}.

About 10\%-30\% of the ETGs imaged in the UV with 
{\it GALEX} show signatures of such rejuvenation episodes, even after excluding 
classical UV-upturn candidates \citep[e.g.][]{Yi05, Donas07, Schawinski07}.
Either accretion events or interactions seem to be at the origin of rejuvenation
episodes. \citet{Rampazzo07} and \citet{Marino09} found evidence of young
 stellar populations in  the nucleus of ETGs showing  shell structures,
signatures mainly produced by a small accretion event 
\citep[see e.g.][and references therein]{Dupraz86, Ebrova09}. 
 Similar results have been obtained by \citet{Jeong09}.

Wide-field UV imaging from  GALEX has enabled, for the first
time, the detection of  young stellar populations in extreme outskirts of galaxies.
 \citet{Thilker10} discovered recent star formation, likely fueled by an accretion episode, in the
outermost portion of the S0 galaxy NGC 404,  a local prototype. 
 The FUV-bright sources are 
concentrated within an HI ring, signature of a merger event  
\citep{delRio04}.  \citet{Marino09} found evidence of
recent star formation ($<$1 Gyr) in the polar-ring early-type galaxy MCG-05-07-01. 
Polar rings are sometimes generated by major merging events. The star formation
in the UV polar ring is also in this case associated with an  HI ring. Recently, 
\citet{Bettoni10}  explored the case of another S0 galaxy, NGC~4262, where they
found an HI and a UV ring. The kinematics of the cold
and warm gas associated with the UV ring is decoupled with that of the galaxy 
stellar population.  The kinematical decoupling indicates that the origin of the outer ring  
is a past major interaction episode undergone by NGC~4262 which is responsible
for the onset of the star formation in the UV-bright ring. 

Presence of ring structures represents a normal phase in the secular evolution 
of S0 galaxies \citep{Buta96}. They are indeed found in about 20-30\% 
 of S0 and spiral galaxies \citep[see e.g.][]{Thilker07, Aguerri09}.
In most cases, rings are associated with non-axisymmetric 
structures, like bars, ovals or triaxial bulges.  \citet{Jeong07} presented 
a  GALEX study of NGC~2974.  This misclassified S0 galaxy shows an  
arm-like structure in UV bands and possesses  a 
bar as deduced from the perturbation of the stellar kinematics 
\citep{Krajnovic05}.  Recently \citet{Salim10} found 
star forming rings, likely of different origin, in 15 out of a sample of 
22 ETGs having extended UV emission.  

If we exclude polar rings produced by minor/major accretions events
and collisional rings, usually rare and recognizable from
the off-centered nucleus, 
secular evolution is at the origin of rings in barred galaxies. 
Nuclear, outer and inner rings correspond to the inner (ILR), outer (OLR) 
Lindblad's and ultra-harmonic dynamical resonances between 
the epicyclic oscillations in the stellar component and the rotation 
of the bar. The gas accumulates in these resonances and can 
induce star formation. Bars either spontaneously or externally triggered, 
can also  drive the gas to the center of  the galaxy, inducing nuclear star formation
\citep{Combes08}. 
The secular evolution driven
by bar instabilities may then represent for a significant fraction of S0
galaxies a  global rejuvenation mechanism whose effects 
extend from the galaxy nucleus to the galaxy outskirts.

 Is the secular evolution induced by bars a viable global
rejuvenation mechanism  in ring S0 galaxies? Is  wet accretion a necessary 
condition to ignite star formation (SF) in ring S0s?
In this  paper we analyze five S0s showing outer rings 
and/or arm-like structures prominent in the  UV bands, 
selected from a larger sample of ETGs  located in low density 
environments.  We describe the sample properties 
 and the observations in Section~2 and 3, respectively.
 We derive the UV and optical photometry
in Section~4.  We use synthetic stellar population models to
derive ages and stellar masses for both the ring and the entire
galaxy in Section~5.  In Section~6 we discuss
the influence of the environment, the HI distribution, the SF properties
and the color of S0s as a whole and  of their ring structures on
the CMD  in the context of galaxy evolution. Finally, we
compare these S0s showing rejuvenation signatures with a larger sample 
of ETGs whose spectral energy distributions (SEDs) are consistent 
with mostly passive evolution. 
  
\section{The Sample properties}
\label{sample}
 We analyze five S0s, namely NGC 1533,  NGC~2962,  NGC~2974,  NGC~4245 and NGC~5636, 
 showing prominent  outer rings (Figure 1) in UV  wide-field imaging.  
 The galaxies  are part of  a  wider sample of  ETGs, which we studied with a
 multi-wavelength approach \citep{Rampazzo05, Annibali06,
Annibali07, Annibali10, Panuzzo10, Marino10a}.
The wider sample includes ETGs which are members of groups of different richness, and show
 presence of an ISM component in at least one of the following bands: IRAS 
100 $\mu$m, X-ray, radio, HI and CO \citep{Roberts91}.
Most  of them have a LINER nucleus  as shown 
by \citet{Annibali10} and \citet{Panuzzo10} from analysis of optical  and  mid-Infrared spectra rispectively.
The  UV bright  features of this sample warrant a different analysis 
are the subject of this work.  All our sample  but NGC~2974 are classified as barred S0s.
NGC~2974 has been erroneously classified as an elliptical  in RC3 
\citep{devau91}, although  an exponential disk is evident both from the optical and UV 
luminosity profiles  \citep[see also][]{Jeong07, Jeong09, Marino10a}.  
In addition,   \citet{Krajnovic05} found, with {\tt SAURON} 
observations \citep{deZeeuw02},  the existence of non-axisymmetric 
perturbations consistent with the presence of inner bar in NGC 2974. 
We conclude that all our systems are barred S0s with outer rings.

All our S0s are located in low density environments since they are members 
of loose groups of galaxies. NGC 1533 is a member of the Dorado group.   
NGC 2962, NGC2974,  NGC 5636 and its elliptical companion NGC 5638,
belong to groups LGG 178, LGG 179, and LGG 386, respectively \citep{Garcia93}.
NGC 4245 is located in the spiral-dominated group USGC-U478 \citep{Ramella02}.

\section{Observations}

Imaging  was obtained with   GALEX  
\citep{Martin05, Morrissey07} in two ultraviolet bands, FUV
($\lambda_{eff}$ = 1539 \AA, $\Delta \lambda$ = 1344 - 1786 \AA) and
 NUV ($\lambda_{eff}$ = 2316 \AA, $\Delta \lambda$ = 1771 - 2831 \AA)
 with a spatial resolution of 4.2/5.3 arcsec in FUV and NUV
 respectively. The observations come partly from the {\it
 GALEX} public archive\footnote{ http://galex.stsci.edu} and  partly from  our
 Cycle~3 program (ID GI3-0087 PI R. Rampazzo). 
 The exposure times range from 1520 sec (NGC 1533) to 2657 sec (NGC
 2974) corresponding to 3$\sigma$ limiting mags of 
$\sim$ 22.6/22.7  in FUV/NUV  or deeper \citep{Bianchi09}.
 In addition to the UV data, we used optical Sloan Digital Sky Survey
(SDSS) imaging in the u [2980-4130 \AA], g [3630-5830 \AA], r [5380-7230
\AA], i [6430-8630 \AA], and z [7730-11230 \AA]  bands, available for
NGC 2962, NGC 4245 and NGC 5636. 

Figure \ref{fig1} shows the  GALEX  FUV/NUV, and   SDSS color 
composite images, and the surface brightness profiles in FUV, NUV 
and r bands.  Inner bars are better visible in the optical image 
(having 3 times higher resolution than UV imaging), while 
outer rings are clearly enhanced in FUV.
NGC~2962 displays an inner ring more prominent in the optical
image, and an outer ring, well defined in the  GALEX 
imaging (see Figure~1).  In all cases  the outer structures are more sharply 
defined in FUV (which traces the youngest populations) than at optical 
wavelength where a mixture of populations is contributing. 
Resolution and depth differences between UV and optical images may partly
contribute to the observed differences in galaxy morphology.  

\section{Photometry and surface brightness profiles}  
 
We used FUV and NUV  GALEX background-subtracted intensity images
to compute integrated photometry of the galaxies. The FUV and NUV
magnitudes (AB system)   within elliptical apertures 
are given in  Table~1 for each galaxy . 
Errors on UV magnitudes were estimated by propagating the Poisson
statistical error on source and background counts. Background counts
were estimated from the sky background image and  high resolution
relative response map provided by the GALEX pipeline in the same
apertures used to compute the galaxy luminosity. In addition to the
statistical error, we added an uncertainty to account for the systematic
uncertainties in the zero-point calibration of 0.05 and 0.03 mag in FUV
and NUV  respectively \citep{Morrissey07}.  

Columns  2,  and 4  in Table~1 provide for each galaxy  
the adopted distance,  and the  major axis of  the optical  $\mu_{25}$ isophote \citep{devau91},  respectively.
R$_{TOT}$ (column 5) is  where the NUV surface brightness is 
3$\sigma$ above the background,  a/b (column 6)  and PA (column 7) in  Table 1 are  the length of the major axis, 
 the major axis/minor axis and the position angle.  Total FUV and NUV  magnitudes in Table 1 (columns 8-9) were computed in elliptical apertures within R$_{TOT}$. 
 
 After registering the SDSS images (corrected frames with the soft bias 
of 1000 counts subtracted) to the corresponding {\it GALEX} NUV images, 
SDSS magnitudes\footnote{We  converted SDSS counts to magnitudes following 
the recipe provided in {\tt http://www.sdss.org/df7/algorithms/fluxcal.html
 \#counts2mag}.} were computed within the same apertures used in 
the UV  (columns 10 --14 of  Table~1).

 Figure~1 (right panels) shows background-subtracted,
 foreground-extinction corrected   UV surface  brightness profiles 
 obtained using the {\tt ELLIPSE} fitting routine in the   
{\tt STSDAS} package of {\tt IRAF} \citep{Jedrzejewski87}. 
 The extinction correction for the foreground component
was performed using the E(B-V) values taken from NED and reported 
in Table 1 (column 3)  assuming Milky Way dust  with R$_v$=3.1 \citep{Cardelli89}. 
The A$_{\lambda}$/E(B-V)  may vary significantly at UV wavelengths 
\citep[see Table 2 of][]{Bianchi11}. However in all our sample galaxies E(B-V) is very small and the 
adopted correction does not influence the results.  
 Rings produce a hump in the  UV-flux profile and are bluer than  the
 body of their respective galaxy. The luminosity of the ring
 structures was measured   integrating the surface  photometry between 
 the inner (R1) and outer (R2) radii reported in Table~1, (col. 5) and 
 indicated in Figure~1 (right panels) with  vertical blue dashed lines. 
 R1 and R2 were chosen from the FUV profiles where rings
 are prominent and well defined. The same R1 and R2 
 have been then used in the NUV and optical bands where rings are
 much weaker. In order to evaluate the ring luminosity we   subtracted 
 the galaxy contribution.  The underlying galaxy profile was
 estimated (blue solid line in Figure~1) by fitting a spline to the 
 luminosity profile excluding the ring portion. 
 Approximately 25\%, 71\%, 30\%, 33\%, and 60\% of the total FUV luminosity 
of NGC 1533, NGC 2972, NGC 2974, NGC 4245 and NGC 5636  comes from the rings. 
In the optical bands, rings are not prominent (at the SDSS depth and S/N) 
and only upper limits were obtained.

 Figure \ref{fig2} shows FUV background-subtracted images 
 with  superimposed the optical D$_{25}$  and the NUV -defined R$_{TOT}$ ellipses obtained 
 as described above. The UV aperture R$_{TOT}$ is in all cases larger than the optical one.
 UV ellipticity and orientation differ from the optical values, 
suggesting that UV young stellar populations in  outer rings are 
rather different than those in the central part of the  respective galaxies. 
For NGC 2962 and NGC 5636 the optical D$_{25}$ ellipse does not include
the extent of the entire galaxy in FUV (figure \ref{fig2}).  

\section{Ages and stellar masses of the galaxy populations}

We  characterize the  epoch and  strength of the rejuvenation episodes in
the S0 sample by estimating  ages and stellar masses  of  the entire galaxies and of their 
outer rings.
We compared the observed spectral energy distributions (SEDs)
with a grid of synthetic stellar population models computed by \citet{Marinolga} with the  GRASIL  
code \citep{Silva98},  which takes into account the effects of the 
age-dependent extinction with young stars being more affected by dust.
We used our grid for  two different star formation histories (SFHs): one 
typical of elliptical galaxies, characterized by a short (1~Gyr) 
intense period of star formation followed by pure passive stellar evolution,  
and one  typical of Spirals with a more prolongued star formation.
SFH parameters are given in Table 5 of  \citet{Marinolga}.  The model grid was computed assuming
 a Salpeter Initial Mass Function (IMF) with a mass range 
between 0.15 and 120 M$_{\odot}$
for Ellipticals and between 0.1$-$100 M$_{\odot}$ for Spirals 
following \citet{Silva98}.  
Our  GRASIL model grid was computed for ages from a few Myr to 13 Gyr  
\citep[][see in particular their Figure 10]{Marinolga}.
The best-fit model obtained by $\chi^2$ fitting of the observed 
SED (FUV to SDSS-z band), and interpolating within the grid, yields the age 
  of the composite population  and the foreground extinction. The  stellar mass is derived by scaling the 
  best-fit model    to the 
  observed SEDs, using the distances in Table 1 and taking into account extinction.
  
In Figure~\ref{sed} we plot  the SEDs within R$_{TOT}$ (full dots) 
of  NGC~2962, NGC~4245, NGC~5636,
 the three sample galaxies for which UV and optical measures are available, 
and of the elliptical NGC 5638, from \citet{Marino10a},  which does not show 
UV-bright ring features.  The solid line represents the best-fit model obtained adopting 
an Elliptical SFH and assuming the foreground extinction given in Table~1
 in addition to the internal extinction accounted 
by {\tt GRASIL} (left panels).   Right panels  show  the best-fit obtained 
 considering an additional  E(B-V) component  as a free parameter in the fitting. 
 By imposing a maximum value of foreground E(B-V) as 
given in Table 1,  the  best-fit model matches well  the optical SED of all S0s but an observed
 FUV excess  indicates an additional   young stellar component.    
Reproducing both the FUV excess and the optical colors with simply passive SFH and additional
reddening (right panels) requires unrealistic high values of
 the resulting E(B-V)  and can be ruled out.  
Such FUV excess is not seen in NGC 5638. The formal  ``best fit"  age of NGC 5638 of  12.6  Gyrs
agrees with the   independent estimate of 9.1$\pm$2.3 Gyrs
obtained by \citet{Annibali07} from Lick  indices in the nuclear region.

Lacking optical data for NGC 1533 and NGC 2974,  we used only
FUV and NUV measures to estimate the age using the same GRASIL grids.
In this case, ages are  largely uncertain since the contribution 
of the young component cannot be separated from the
passively evolving population.   
  
For estimating ages and stellar masses of outer  ring structures we are compelled
to use only FUV and NUV measures since only upper limits 
can be derived from the SDSS images. Ring FUV and NUV 
magnitudes were compared both with GRASIL and  Single-burst Stellar Population (SSP)   
 \citep[e.g.][]{Bianchi11}  models.   
 Comparing the FUV - NUV color with  SSP,  Elliptical and Spiral SFH model color yields markedly different  
   ring ages (column 4 in Table~2 ). We derive very young ages ($\la$ 200 Myr)
adopting a SSP SFH  and ages of  $\sim$ 1Gyr with an Elliptical SFH. The ages
derived assuming Spiral SFH appear  older than those of the entire galaxy,
ruling out the assumption of continuous  SF for the ring population.
The most plausible SFH for the population of the outer UV-emitting rings
appears to be SSP-like. An exponentially decaying SFH, would give similar results as the SSP models 
in  \citep[e.g][]{Efremova11}.
The very different ages obtained when assuming different SFHs and the comparison of
 observed FUV-NUV colors with  model colors in Figure 4, underscore the
importance of obtaining significant measurements of the outer ring
structures at longer wavelengths, which we will pursue with
deep optical imaging.  

The stellar mass of the outer rings is about 1-4\%, $<1$\%, 
1\% and 5-8\% of the total stellar mass of NGC 1533, NGC 2962, NGC 2974, 
NGC 4245, and NGC 5636, respectively.  
The estimate of stellar mass is affected by uncertainties in the
photometry, extinction, distance and derived ages.
The  provided  errors take into account only the photometric error, 
 but not  the larger indetermination  which may result  from the assumption of a given SFH.

In column 6 of Table~2 we also report the stellar mass surface density 
(average over the photometry aperture). 

\section{Discussion}

\subsection{The Environment} 

Possible signatures  of either  recent accretion events or of  ongoing interactions
are reported in the literature for three of our S0s. 
\citet{deGraaff07} describes NGC~1533 as a late stage of 
transition from a barred spiral to a barred S0 galaxy. 
\citet{Grossi09} and \citet{Werk10} found HI tails in NGC~2962
and NGC~1533, respectively (see next subsection).
\citet{Tal09} report the presence of a shell system surrounding NGC 2974.   

In order to evaluate an interaction probability  
the $f$ perturbation parameter 
 has been computed.
This parameter,  defined  in \citet{Varela04} as:

$$ f=log(F_{ext}/F_{int})= 3 log(R/D_p)+0.4\times(m_G-m_p) $$

is  the logarithmic ratio between inner ($F_{int}$) and tidal forces ($F_{ext}$)
acting upon the galaxy by a possible perturber (where $R$ is the size of the 
galaxy, $D_p$ the projected distance between the galaxy and the perturber,
$m_G$ and $m_p$ the apparent magnitudes of the primary and the perturber
galaxies, respectively). \citet{Varela04} show that ``unperturbed'' galaxies 
(which they call ``isolated'') have $f  \leq -4.5$. 

To compute $f$ we consider as possible perturbers the galaxies taken from {\tt HYPERLEDA} 
that have $M_B$ brighter than $-12$mag, $D_{25}$ corresponding to more than 
2 kpc, $\Delta V \leq 500$ km~s$^{-1}$,  with the primary and projected 
angular separation $\leq 5^{o}$.  Within the above conditions our barred S0s have 
$-6.4 (NGC 2974) \leq f \leq -0.74 (NGC 5636)$.  At face value, $f$ indicates in
all cases but NGC 2974,  that our UV-ringed S0s  
currently inhabit sparsely populated environments. In such environments 
 strong interaction episodes can take place, since
the velocity dispersion of the galaxies 
is comparable to the galaxy stellar velocity dispersion.   HI tidal tails 
and shells,  often revealed in ETGs located in low density environments, are signatures  of recent dynamical perturbations
\citep[][and references therein]{Reduzzi96, Rampazzo06, Rampazzo07}.

\subsection{Star formation and HI distribution}
 
FUV and  HI flux distributions usually correlate in star-forming regions
\citep[e.g.][]{Neff05, Thilker05, Xu05}. HI masses, computed from the
flux\footnote{taken from  NED} 
with  the relation M$_{HI}$=2.36$\times$10$^5$ d$^2$ S$_{HI}$,   
where $d$ is the distance in Mpc and S$_{HI}$ is the integrated flux in Jy km~sec$^{-1}$, are
reported in column 7 of Table 2  together with the respective area when available. 
In NGC~1533, \citet{Weber03} report  HI arcs around the
galaxy, spanning from 2\arcmin\ and 11.7\arcmin\  from the optical  center of the
galaxy,  i.e. outside of   the FUV ring, with a mass of 7 $\times$ 10$^9$ M$_\odot$, 
and conclude that the HI is most likely the remnant of a tidally destroyed galaxy. 

 \citet{Burstein87}  found that most of the gas  may be in a ring approximately coincident
 with the outer disk of NGC 2962. They  measure a flux of 1.3 $\pm$ 0.2 Jy km~sec$^{-1}$
 within a radius less than 2.6\arcmin\  that is 
 comparable to R$_{TOT}$.  Adopting this HI flux, the M$_{HI}$ over stellar mass ratio is between $\sim$   0.4 
 (assuming E(B-V)=0.07 mag) and $\sim$ 0.8 (assuming E(B-V)=0.67 mag). 
 The HI flux associated with NGC 2962 as measured by \citet[][4.22 $\pm$ 0.12 Jy km~sec$^{-1}$]{Grossi09}   and 
 \citet[][5.3 $\pm$ 2.5 Jy km~sec$^{-1}$]{Wa86},  is significantly 
 greater than the value measured by  \citet{Burstein87}.
\citet{Grossi09} found that the HI emission of NGC 2962 appears to be connected to that
of the galaxy SDSSJ094056.3+050240.5 located at a projected distance of
$\sim$ 8\arcmin\ from NGC 2962.

\citet{Kim88} measured a HI flux associated with NGC 2974, equal to the value we adopt, in an area of $\sim$ 
4\arcmin $\times$ 2.4\arcmin, about 20\% larger than the optical extent of the
galaxy and with a distribution aligned with the optical isophotes.  From the 
{\tt NED} data-base, NGC 4245 is the most HI deficient  in the sample with a HI
flux of 0.77 $\pm$  0.14 Jy km~sec$^{-1}$.

 In sum, HI emission has been detected in all our S0 galaxies with UV-bright 
 rings, and in one case (NGC~1533) in its nearby environment. 
\citet{Oosterloo10}   
found that HI is detected,  with a detection
limit of a few times 10$^6$ M$_\odot$ in $\sim$ 2/3 of the field ETGs
while $<10$\% of   Virgo members have HI detections. 
Basically all fast rotating galaxies, 
like typical S0s, have a HI disk. 
The HI detection in our S0s, 
members of groups, is not an exception. 
 Furthermore,  \citet{Oosterloo10} notice that field S0s are ``rejuvenated''
 with respect to  their cluster counterparts.  This is also the case of our 
barred ring S0s: secular  evolution,  accretion of fresh gas,  or both, taking place  in 
low density environments, may feed secondary star formation events.    

\subsection{The UV-optical CMD} 

In Figure~\ref{ss} we plot  the UV-optical CMD of the ring structures,
of the galaxy central part (empty circles) and of the entire galaxies (filled circles)
for both our sample  and a comparison sample of 13 ETGs without 
outer rings from \citet{Marino10a}.  We also overplot the \citet{Yi05} fit   
 to the `red  sequence' in ETGs  (solid line).  We recall that for the ring
structures only an upper limit to the optical flux was obtained.
The ETGs in  \citet{Marino10a}   have a large spread in the luminosity-weighted ages
 as estimated by  \citet{Annibali07} from line strength indices.
 The youngest ETG is NGC~3489,  with an estimated age of 1.7$\pm$0.1 Gyr for the nuclear 
 population, while the oldest is NGC~5363, with an age of 12$\pm$2.3 Gyrs. 
 Such ETGs lie in the red sequence 
 irrespective of the luminosity weighted age of their nucleus. 
\citet{Annibali07} do not provide a luminosity weighted age for the
nucleus of NGC~2962. The (NUV-$r$) color of the central portion of this
S0 is redder than the total galaxy color  although it includes the ring
structure. 
The  (NUV-$r$)  central and total colors of  NGC~4245 and NGC~5636
progressively detach from the red sequence and place the
galaxies in the so called green valley,  where  evolving galaxies are located
\citep{Salim07}. Ring structures (triangles in Figure~\ref{ss}) lie in the  
blue sequence locus. 
 
 The location of the rings and of the entire galaxies appear well separated also in the 
 (FUV-NUV) versus (NUV-$r$) plane \citep[see][their figure 3]{Marino10c}.

\section{Summary and Conclusions} 
 
 We derived the properties of the stellar populations in outer ring
 structures and in the main galaxy in a sample of barred S0 galaxies 
 from model analysis of  UV and optical photometry. 
 We find evidence of recent star formation in the rings whose
 stellar  populations  are younger than the mean galaxy population. 
 The stellar masses of  the outer rings are only a few percent of the galaxy
total stellar mass while the luminosity accounts for up to 70\% of the total 
FUV flux.
  
\begin{figure}
\centering
\includegraphics[width=13.0cm]{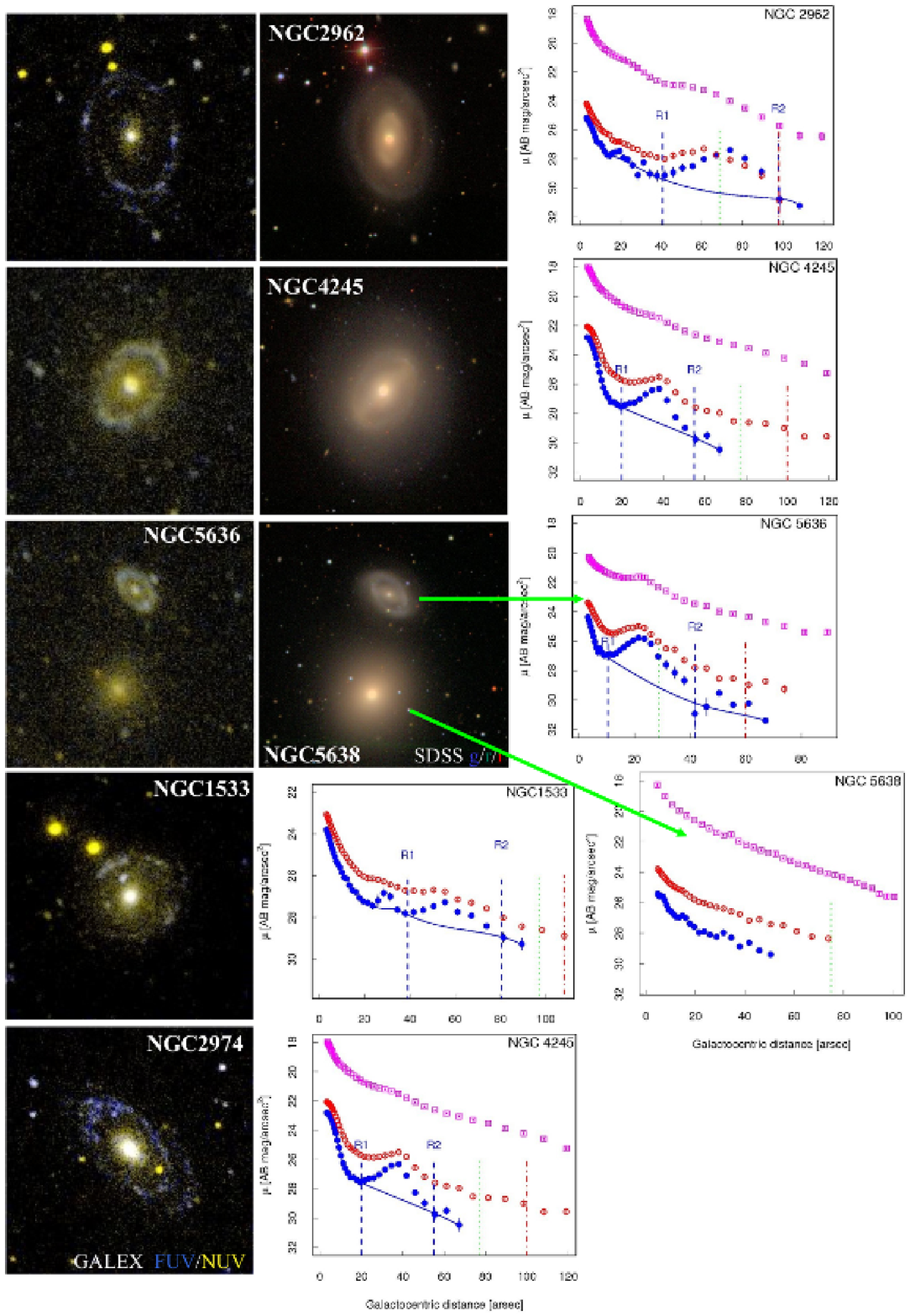}
\caption{Composite  FUV (blue) and NUV (yellow)  {\it GALEX}
images of all targets and  optical SDSS-g/r/i (blue/green/red) images of NGC 2962, 
NGC 4245, and NGC 5636.  Luminosity profiles along the semi-major axis 
are shown in FUV (blue filled dots), NUV (red empty
points) and SDSS-r bands (magenta squares).
The vertical lines  indicate the ring inner and outer radii (blue), optical 
D$_{25}$ (green), and NUV-defined R$_{TOT}$ (red: see text). 
For comparison we also plot  the luminosity profile of the 
elliptical galaxy NGC 5638 \citep{Marino10a},  which does not show outer rings.}
\label{fig1}
\end{figure}

\begin{figure*}
\includegraphics[width=6.cm,height=6.1cm]{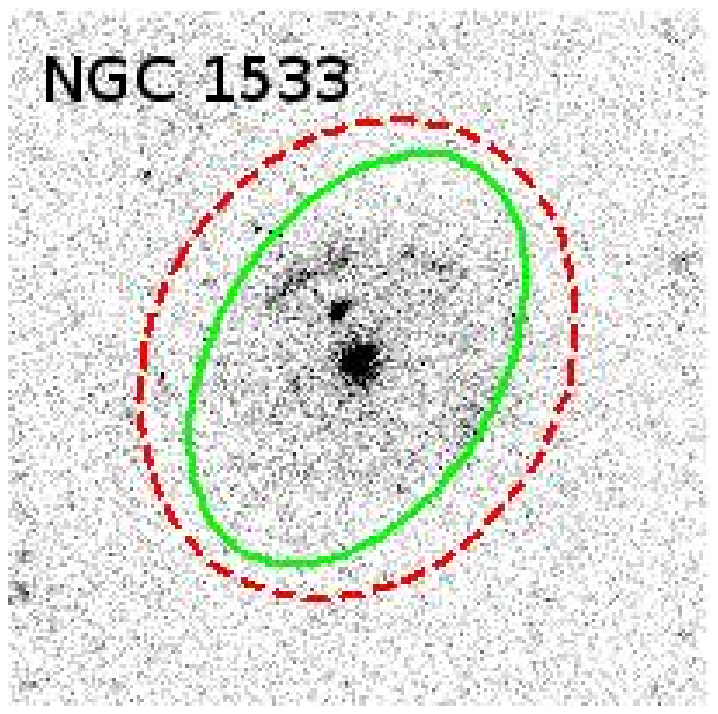}  
\includegraphics[width=6cm]{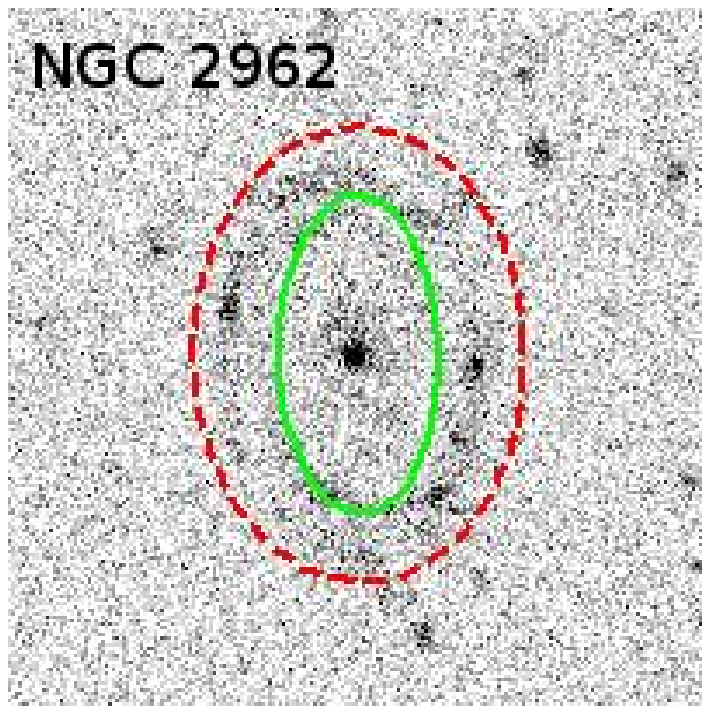}
 
\includegraphics[width=6cm,height=6.cm]{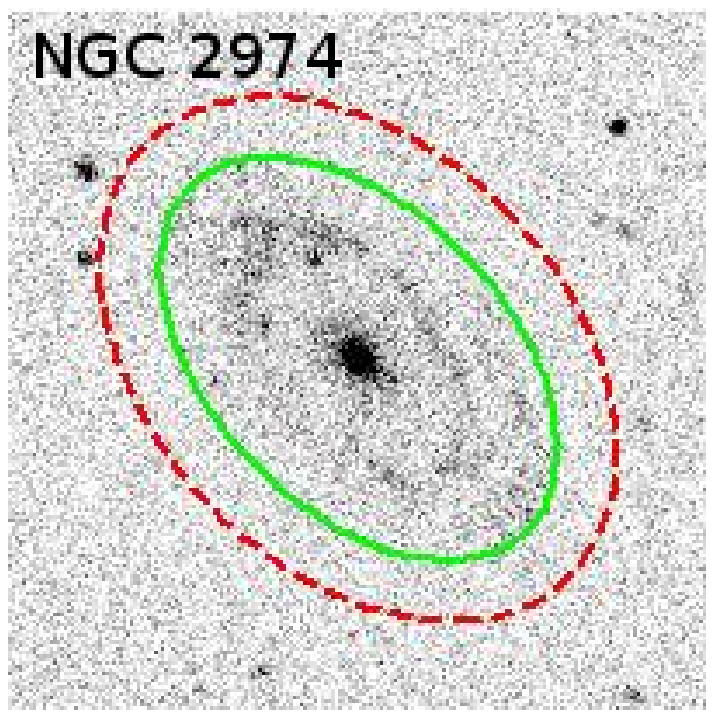}
\includegraphics[width=6cm,height=6.cm]{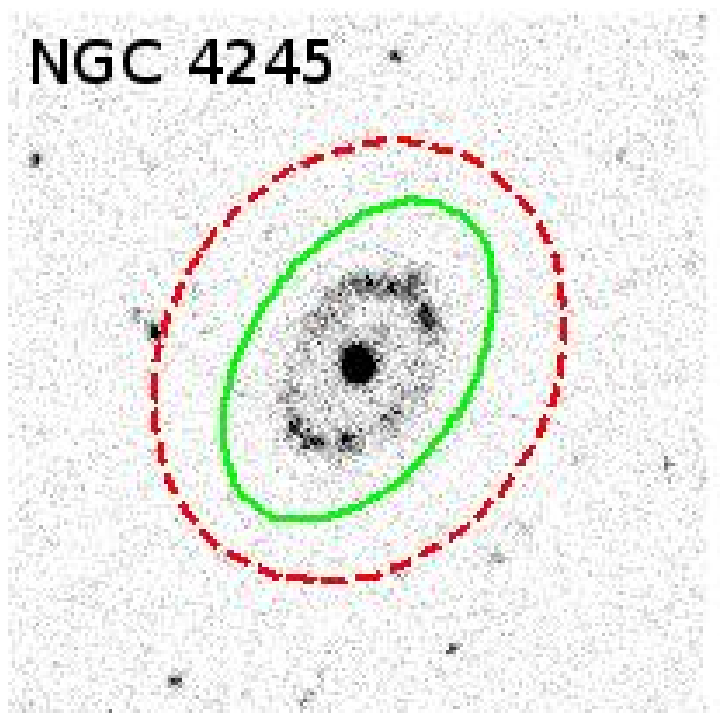}
 
\includegraphics[width=6cm]{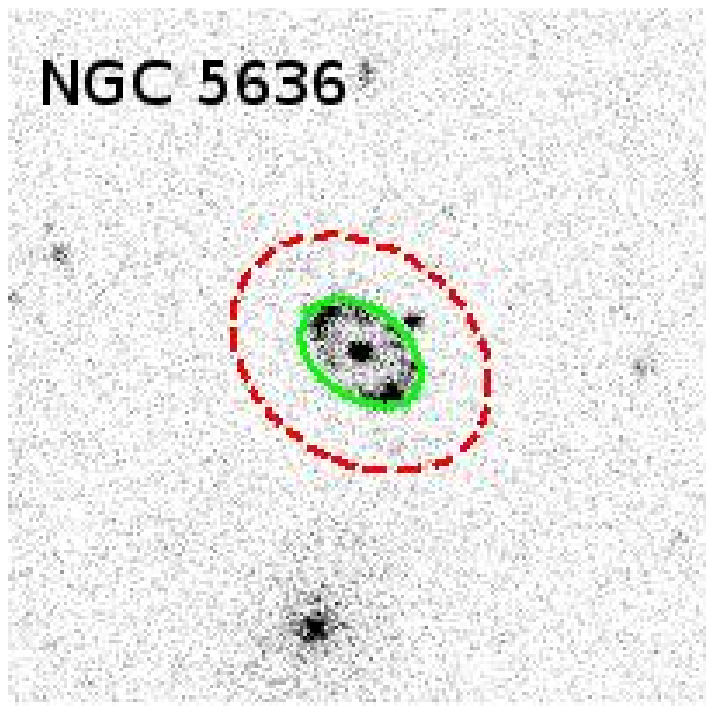}
\caption{ 5\arcmin\ $\times$ 5\arcmin\ {\it GALEX} FUV 
background-subtracted images of the sample with  
the optical D$_{25}$ ellipse (green solid line) superimposed. Red  
dashed lines indicate the R$_{TOT}$ ellipse,  
defined from NUV light profiles (see text). 
For NGC 2962   and NGC 5636  the
optical diameter does not include the extent 
of the entire FUV emission.} 
\label{fig2}
\end{figure*}

\begin{figure}
\centering\includegraphics[width=14.5cm]{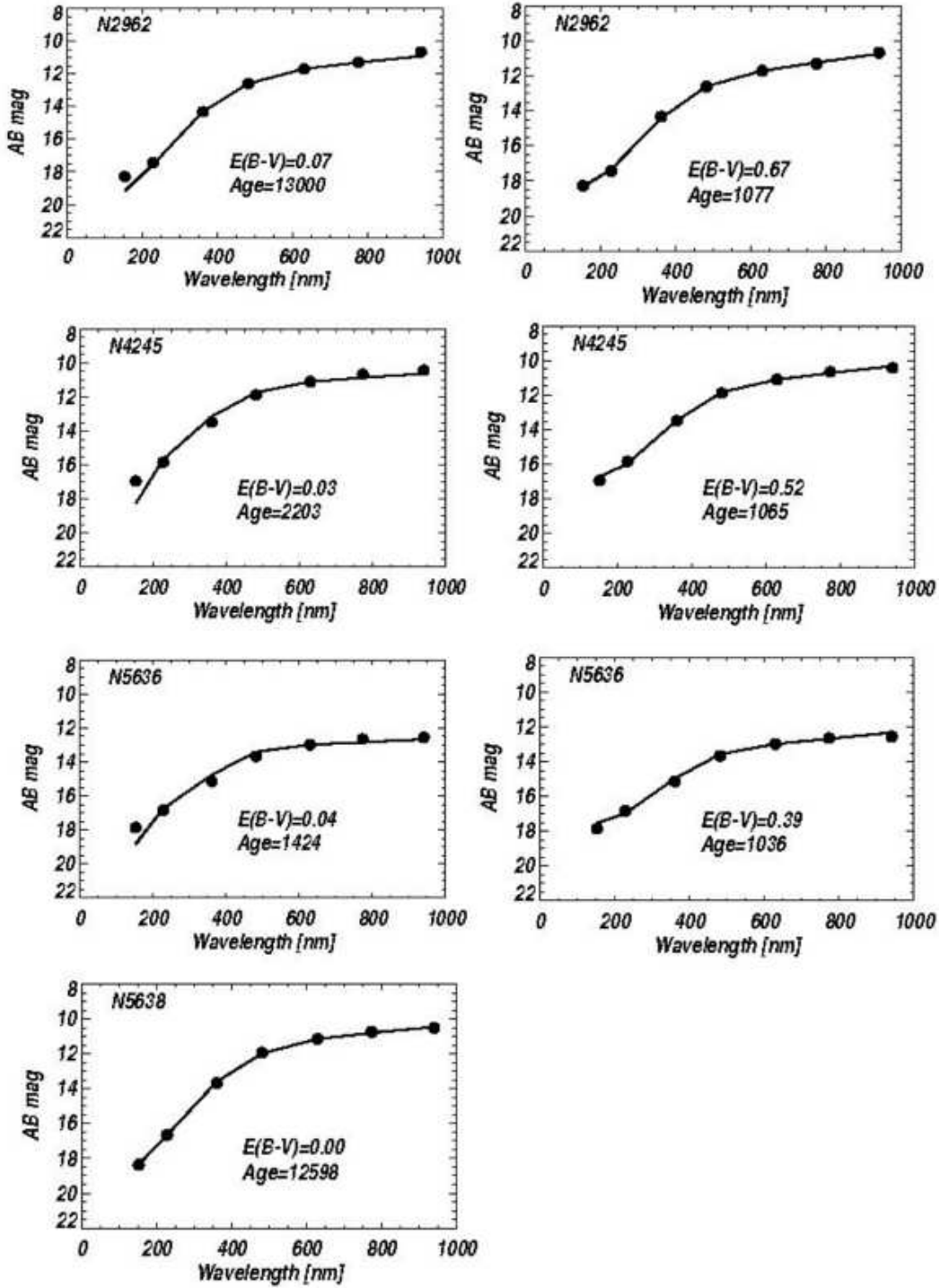}
\vspace{-0.6cm}
\caption{FUV to near-infrared SEDs  (dots), 
measured within R$_{TOT}$ given in Table~1, 
with over-plotted the best-fit
models   (lines) obtained assuming  an Elliptical SFH  and either the foreground extinction 
given  in  Table 1 
(left panels, showing a FUV excess in the data) or leaving  E(B-V)  
as a free parameter (right panels). Ages are in Myrs. 
In the bottom left panel
we plot the SED and the best fit model of the elliptical NGC 5638.}
\label{sed}
\end{figure}

 The UV, star-forming rings in NGC 4245, NGC~5636 and
NGC~1533 are located at the end of the bar structure. The 
optical and UV images, shown in Figure~\ref{fig1},
suggest that NGC 2962 has two rings, one at the end of the bar,  and one more external, 
 prominent  in the UV {\it GALEX} 
images, showing arm-like structures.  The ring and the outer features detected
in NGC~2974 are reminiscent of  NGC 2962, which is seen
more face-on. 
  
All S0s in our sample show evidence of a bar in optical imaging. 
UV-bright  outer rings are  likely related to the bar perturbation. 
At the same time, there are several indications that NGC 1533,
NGC 2962 and NGC~2974 may have been re-fueled of fresh gas by either
nearby gas-rich companions (NGC 1553 and NGC 2962) or  by a minor 
merger event, considering the presence of shells (NGC 2974). 
 Recently,  \citet{Serra10} have shown that a large fraction of ETGs have
continued their assembly over the past few Gyrs in the presence of
a mass of cold gas of the order of 10\% of the galaxy stellar mass:
a material that is now observable as HI and young stars. Their study
include both NGC~2974 (to which they associate a central luminosity weighted
age of 3.0$^{+1.0}_{-0.6}$ Gyr) and NGC~5638 (8.1$^{+0.8}_{-1.2}$ Gyr).

We suggest that rings in our barred S0 sample are an effect of the secular 
(inner) evolution following the formation of a bar likely aided by fresh gas 
accreted from gas-rich perturbers. 
The tumbling bar drives gas inwards and the gas accumulates  at the 
Lindblads resonances \citep{Buta96} where star formation
may take place. 
In all S0s but NGC~4245,  
HI is still detected  outside the galaxy rings. 
Rebuilding  the  secular evolution scenario of the S0 NGC 4570, 
 \citet{vandenbosch98} 
concluded that the mass of the  outer rings does not require a 
large gas supply,  and may be even consistent with these structures having 
formed from mass loss by the older population of stars. This could  apply also to  
NGC 4245 which is HI poor \citep{Ga94}.  

\citet{Kannappan09} identify a population of morphologically defined
E/S0 galaxies lying on the locus of late-type galaxies in color-stellar mass space
$-$the blue sequence$-$ at the present epoch. In addition they
 noticed that ``bars, rings and dust are evident in both 
 red$-$ and blue$-$sequences of E/S0s". The color and the masses computed 
 for our  barred S0s with UV bright rings locate
them in the so called  green valley  \citep{Salim07}. 
We are likely catching them in the evolving path following their rejuvenation produced
by the interplay between secular and externally driven evolution (in-flight re-fueling, 
small wet accretion), evolving back to the E/S0 red$-$sequence.

\begin{acknowledgments}

A.M.  acknowledges support from the Italian Scientist and Scolar of
North America Foundation (ISSNAF) through an ISSNAF fellowship in Space
Physics and Engineering, sponsored by Thales Alenia Space. 
 L.B. acknowledges support from NASA grants NNX10AM36G and NNX07AP08G.
RR acknowledges financial support from the agreement ASI-INAF I/009/10/0.
 GALEX  is a NASA Small Explorer, launched in April 2003. 
GALEX is operated for NASA by California Institute of Technology under
NASA contract  NAS-98034. The GALEX data presented in this paper were obtained 
from the Multimission Archive at the Space Telescope Science Institute (MAST). 
STScI is operated by the Association of Universities for Research in Astronomy, 
Inc., under NASA contract NAS5-26555. Support for MAST for non-HST data is 
provided by the NASA Office of Space Science via grant NNX09AF08G and by 
other grants and contracts.  {\tt IRAF} is distributed by the National
Optical Astronomy Observatories, which are operated by the Association
of Universities for Research in Astronomy, Inc., under cooperative
agreement with the National Science Foundation. This research has made
use of  SAOImage DS9, developed by Smithsonian Astrophysical Observatory
and of the NASA/IPAC Extragalactic Database (NED) which is operated by
the Jet Propulsion Laboratory, California Institute of Technology, under
contract with the National Aeronautics and Space Administration. We
acknowledge the usage of the HyperLeda database
(http://leda.univ-lyon1.fr).
\end{acknowledgments}

{\it Facilities:} {\it GALEX}, Sloan

\bibliographystyle{apj}
\bibliography{Rings_rev1}
 
\newpage

\begin{figure}
\centering
\includegraphics[width=14.5cm]{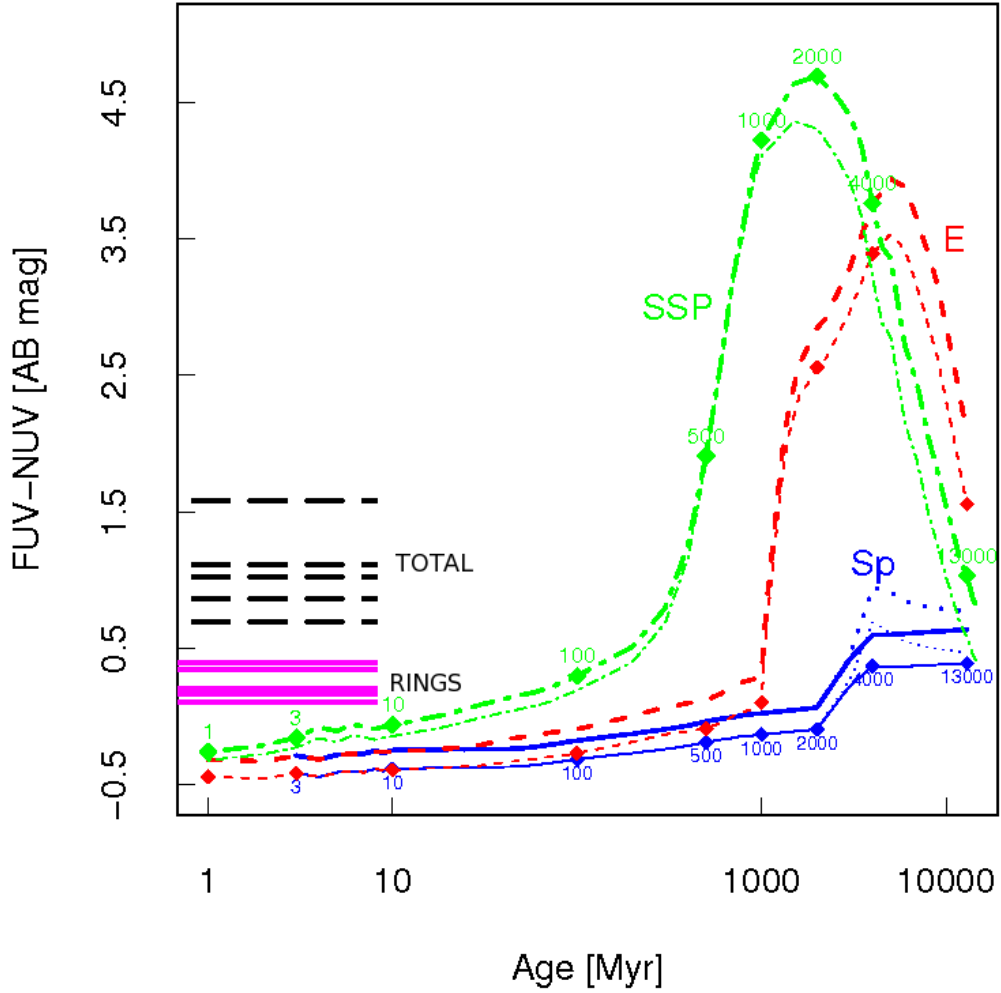}
\caption{FUV-NUV  model color vs. age   for  SSP  (green dot-dashed lines) and
  Spiral (blue solid/dotted lines
for inclination $i=0^{\circ}/90^{\circ}$) and Elliptical (red dashed lines) SFHs. Thin lines are
models with internal extinction but no foreground extinction;
 thick lines are for models  reddened with an additional foreground extinction of E(B-V)=0.5 mag.  Measured
colors of the rings (horizontal magenta lines)  and  of the entire galaxies (black
dashed lines) are also shown. Age labels  are in Myrs (diamonds).}
\label{age}
\end{figure}

 \begin{figure} 
 \centering  
 \includegraphics[width=16cm,angle=-90]{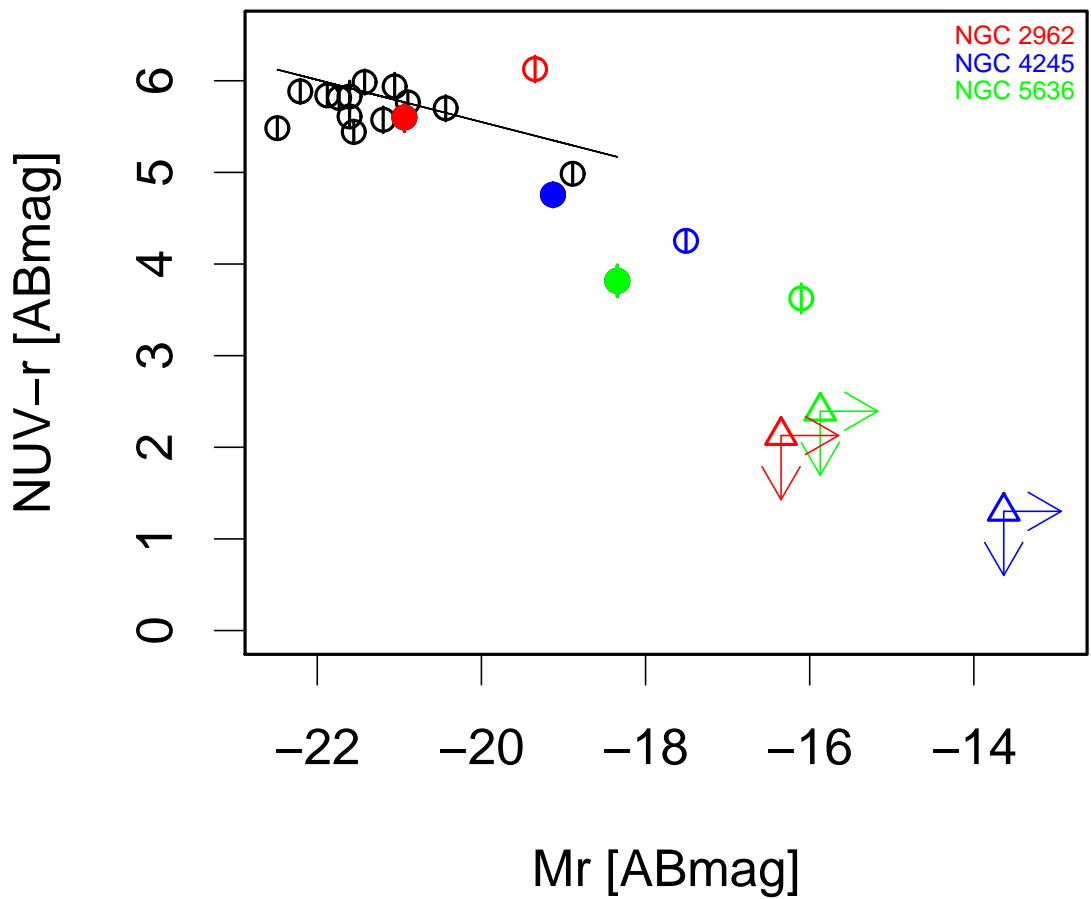}
  \caption{M$_r$ vs (NUV- r) measurements of the ring structures 
   (triangles), of the central galaxy portion ($\leq R_1$; empty circles) and of the total galaxy
   (within R$_{TOT}$;  filled circles) for NGC 2962 (red), NGC 5636 (green) and NGC 4245
   (blue).  
   Black circles are  the measurements  within the optical D$_{25}$ of
   13 ETGs with no evidence of   ring structures from \citet{Marino10a}.  
   We also overplot the \citet{Yi05} fit  to 
    the red galaxy sequence (solid line).
   NGC 2962   is  in the red sequence  
   while ring colors  lie on the blue sequence.  
   NGC 4245 and NGC 5636 lie in the  green valley  \citep[e.g.][]{Salim07}. }
   \label{ss}
 \end{figure}
 
 \begin{sidewaystable}[h]
 \vspace{-8cm}
 \scriptsize{
 \caption{The sample photometry.} 
\begin{tabular}{llllllllllllll}
 \hline\hline
Name &  Dist.   & E(B-V)$^a$&D25  &R$_{TOT}$ [$^{\arcsec}] $ & a/b$^b$ & PA$^c$ &  FUV  & NUV &  u & g & r& i& z \\
     & {[Mpc]}  &[mag]  & [$^{\arcsec}$]     &R1-R2 [$^{\arcsec}]$   &  &[deg] &[AB mag]  & [AB mag]      &[AB mag]    & [AB mag] &[AB mag] &[AB mag] &   \\
\hline
NGC~1533 &  13.4  & 0.015&97.2 &108& 1.2& 149&  16.90$\pm$0.09 & 16.04$\pm$0.05 \\
NGC~1533$^R$ &    &      & &39--80   & &&18.38$\pm$0.11   &18.27$\pm$0.10    \\
\hline
NGC~2962 &  30.6  & 0.058& 68.4 &98 & 1.4 & 1 & 17.78$\pm$0.15 & 17.09$\pm$0.10 &14.19$\pm$0.11 & 12.32$\pm$0.02 & 11.49$\pm$ 0.02 &11.17$\pm$0.02& 10.53$\pm$0.04\\
NGC~2962$^R$ &  & &   &  41--98 & & &18.15$\pm$0.27 & 17.98$\pm$0.31 &\\ 
\hline		     
NGC~2974 &  28.5 & 0.054&104.4 &131 & 1.4& 44  & 17.78$\pm$0.09 & 16.2$\pm$0.05  \\
NGC 2974$^R$ &   &    &  & 35--72 &        && 19.07$\pm$0.15    &   18.86$\pm$0.09    \\
\hline
NGC~4245 &  11.4  & 0.021&76.8 &100 & 1.2  & 146 & 16.95$\pm$0.07& 15.84$\pm$0.04& 13.49$\pm$0.06& 11.88$\pm$0.02& 11.10$\pm$0.02& 10.65$\pm$0.02& 10.42$\pm$0.03  \\
NGC~4245$^R$ &    &    &   &20--55 &       & &18.15$\pm$0.14& 17.80$\pm$0.09   \\
\hline
NGC~5636 & 22.2   & 0.033&28.8& 60& 1.3 & 40 &  17.86$\pm$0.14& 16.84$\pm$0.06& 15.14$\pm$0.13& 13.66$\pm$0.03&12.98$\pm$0.03& 12.65$\pm$ 0.034 &12.56$\pm$0.09  \\
NGC~5636$^R$ & &   &       &10--42  & & &18.41$\pm$0.07& 18.01$\pm$0.11  \\
\hline 
 \end{tabular}

Note: All magnitudes are in the AB system and are uncorrected for Galactic extinction.

$^a$ From  NED

$^b$ a (=R$_{TOT}$) and b are the semi-major and semi-minor axes from NUV light profiles.  Total
magnitudes in columns 8-14 are calculated within the ellipse defined by these axes.

$^c$ From Hyperleda

$^R$Ring luminosity between R1 and R2, defined from the FUV images (Figure 1).
}
\label{tab1}
\end{sidewaystable}

\begin{table*}
\scriptsize{
\caption{Ages and stellar masses of the ring structures and of the entire galaxies.} 
\begin{tabular}{llllccclc}
\hline\hline

Name &    Region & SFH & & Age  & Mass$_{\star}$ & Mass$_{\star}$/Area &M$_{HI}$ (Area)\\
     &         & assumed & &[yrs] & [10$^6$ M$_\odot$] &[10$^6$ M$_\odot$]/kpc$^2$ & [10$^6$ M$_\odot$] (kpc$^2$) \\
\hline
NGC~1533  & Ring & SSP	& $^a$	&   0.7$_{-0.4}^{+0.6}\times 10^8$ &  0.9$_{-0.6}^{+1.1}$ & 0.02 \\

          &  & E        & $^a$	&   10.0$_{-4.0}^{+0.2}\times 10^8$ &  6.4$_{-8.9}^{+1.3}$  & 0.1\\
	  
          &  & Sp	& $^a$	&   2.6$_{-0.5}^{+0.5}\times 10^9$ &  46.6$_{-31.4}^{+83.9}$ &0.9   \\
	  
          & $\leq R_{TOT}$& E &     $^a$ &   11.0$_{-0.1}^{+0.3}\times 10^8$ &  175$_{-37.9}^{+46.1}$ & &3700 (?)    \\
\hline 
NGC~2962  & Ring & SSP   & $^a$     &  0.9$_{-0.5}^{+0.7}\times 10^8$ &   0.4$_{-0.3 }^{+0.5}$ &0.001 \\

          &   & E        & $^a$       &  10.1$_{-2.6}^{+0.1}\times 10^8$ &  2.2$_{-4.4}^{+0.4}$  & 0.01 \\
	  
  &   & Sp               & $^a$        & 2.7$_{-0.4}^{+0.4}\times 10^9$ &  20.3$_{-12.6}^{+26.3}$ & 0.05  \\
  
  & $\leq R_{TOT}$& E &   $^b$    & 13:$\times 10^9$(E(B-V)=0.07)  & 690: & &287 ($\sim$1700) \\
  &      &   &   $^b$    & 1.1:$\times 10^9$(E(B-V)=0.67)  &347: & & \\
\hline
NGC~2974   & Ring & SSP	 &  $^a$   & 1.1$_{-0.7}^{+0.9}\times 10^8$ &   0.2$_{-0.2}^{+0.4}$ & 0.002\\

           &  & E        &  $^a$  &    10.1$_{-2.6}^{+0.2}\times 10^8$ & 1.1$_{-2.0}^{+0.3}$ & 0.01	\\
	   
   &  & Sp               & $^a$     & 2.8$_{-0.5}^{+0.4}\times 10^9$ &		       12.6$_{-9.0}^{+16.7}$ &0.1 \\
   
   & $\leq R_{TOT}$& E &         $^a$   & 12.6$_{-0.2}^{+0.3}\times 10^8$  &152$_{-26.2}^{+31.5}$ && 1150 ($\sim$700)\\
\hline 
NGC~4245 &  Ring & SSP  &  $^a$   &	1.8$_{-0.8}^{+0.6}\times 10^8$  &  6.1$_{-3.6}^{+5.1}$ & 0.3\\   
     
         &   & E      &  $^a$   &    10.3$_{-0.2}^{+0.2}\times 10^8$ &  16.3$_{-3.0}^{+3.8}$ &0.8   \\
	 
 &   & Sp             &  $^a$   &    3.7$_{-0.8}^{NA}\times 10^9$ &  356.3$_{-204.4}^{NA}$ &18.4\\
 
 &  $\leq R_{TOT}$ & E  & $^b$ & 2.2:$\times 10^9$(E(B-V)=0.03) & 1844:  & &23  (?)\\
 &        &    &  $^b$ & 1.1:$\times 10^9$ (E(B-V)=0.52) & 2929: & &\\
 \hline
NGC~5636 &  Ring & SSP &  $^a$ &  2.0$_{-0.6}^{+0.5}\times 10^8$ &  1.8$_{-0.9}^{+0.8}$ & 0.04 \\

         & & E         &  $^a$  &	  10.3$_{-0.1}^{+0.2} \times 10^8$ &  3.9$_{-0.6}^{+0.7}$ & 0.1 \\
	 
 &         & Sp        & $^a$     &	    4.0$_{-0.8}^{NA}\times 10^9$ & 97.9$_{-39.0}^{NA}$ &2.2 \\
 
 &  $\leq R_{TOT}$ & E &        $^b$   &1.4:$\times 10^9$(E(B-V)=0.04) & 57: & &150 (?)\\
 &        &   &        $^b$   &1.0:$\times 10^9$(E(B-V)=0.39 & 99: & &\\
\hline 
NGC5638   & Total & E&$^b$ &12.6:$\times 10^9$(E(B-V)=0)     & 1076: && \\
\hline
\end{tabular}

 $^a$from FUV and NUV only. 
 
 $^b$From  SED fits (FUV to near-infrared). The two values of  age and mass are the results of the  best-fit models
   obtained 1) using  the  foreground extinction values from Table 1 and 2) treating E(B-V) 
  as a free parameter (see Section 5).  Results with an additional  E(B-V)  are unrealistic as discussed in section 5
  and are shown `ab absurdo', supporting  the  view that UV light traces a distinct, younger population superimposed
  to a passively-evolving stellar population in these objects.  
  
   }

\end{table*}

\end{document}